\documentclass[aps,twocolumn,amsmath,amssymb,preprintnumbers]{revtex4}
\usepackage{amsmath} \usepackage{amsfonts} \usepackage{amssymb}
\usepackage{bbm}
\usepackage{epsfig}
\usepackage{graphics}
\usepackage{graphicx}
\newcommand{\be}{\begin{equation}}
\newcommand{\ee}{\end{equation}}
\newcommand{\ba}{\begin{eqnarray}}
\newcommand{\ea}{\end{eqnarray}}

\begin{document}

\title[ ]{Where to look for solving the gauge hierarchy problem?}

\author{C. Wetterich}
\affiliation{Institut  f\"ur Theoretische Physik\\
Universit\"at Heidelberg\\
Philosophenweg 16, D-69120 Heidelberg}

\begin{abstract}
A mass of the Higgs boson close to 126 GeV may give a hint that the standard model of particle physics is valid up to the Planck scale. We discuss perspectives for the solution of the gauge hierarchy problem at high scales. Scenarios with an ultraviolet fixed point have predicted a Higgs boson mass very close to 126 GeV if the fixed point value of the quartic scalar coupling is small. In this case the top quark pole mass should be close to 172 GeV. 
\end{abstract}

\maketitle

The ATLAS and CMS collaborations at the LHC have announced evidence for a mass of the Higgs boson in the range of 125-126 GeV \cite{AT}, \cite{CMS}.  A mass in the vicinity of 126 GeV has been predicted \cite{SW} within models of non-perturbative  renormalizability of gravity \cite{We,Re} which lead to a small value of the quartic scalar coupling near the Planck scale. In this note we ask if a Higgs boson mass around 126 GeV, if confirmed,  could give hints for the energy scale where a possible solution of the gauge hierarchy problem could be found.

\medskip\noindent
{\em Fine tuning problem and anomalous mass dimension.}

The Fermi scale of weak interactions, $\langle\varphi\rangle=\varphi_0=175$ GeV, is more than sixteen orders of magnitude smaller than the Planck scale in gravity, $M=(8\pi G_N)^{-\frac{1}{2}}=2.4\times 10^{18}$ GeV.  Within any unified theory of all interactions the small ratio $\varphi_0/M$ calls for an explanation - this is the gauge hierarchy problem \cite{Gel}. There is widespread belief that the solution of this problem  should be found at energy scales not too far from the Fermi scale. Often this is motivated by the so-called fine tuning problem that states that in a unified model the parameters have to be tuned with high precision order by order in perturbation theory and that such a situation is not natural. Supersymmetry or the absence of a fundamental scalar as in technicolor can avoid this fine tuning.

It has been shown \cite{CW1} long ago, however, that the need of fine tuning in every order is purely a shortcoming of the perturbative expansion series. It is absent in renormalization group improved perturbation theory or within functional renormalization \cite{CWFRG}. This can be seen most easily in a setting where the Higgs doublet $\varphi$ is supplemented by a singlet scalar field $\chi$ whose expectation value is responsible for the Planck mass. (We consider here for simplicity a single field $\chi$ - the generalization to several fields being straightforward.) The relevant terms in the effective potential for $\varphi$ and $\chi$ are
\be\label{1}
U=\frac{1}{2}\lambda(\varphi^\dagger\varphi)^2+\gamma(\varphi^\dagger\varphi)\chi^2+U_\chi(\chi),
\ee
with dimensionless couplings $\lambda$ and $\gamma$. (Terms $m^2\varphi^\dagger\varphi$ or $\nu \varphi^\dagger\varphi\chi$ can be absorbed by a redefinition of $\chi$.) We choose conventions for $\chi$ such that its expectation value equals the reduced Planck mass, $\langle\chi\rangle=M$. Electroweak symmetry breaking occurs for $\gamma < 0$, and the gauge hierarchy needs the explanation of a tiny absolute value of $\gamma$ (for real positive $\varphi_0$)
\be\label{1A}
\frac{\varphi_0}{M}=\sqrt{-\frac{\gamma}{\lambda}}.
\ee

The running of $\lambda$ and $\gamma$ with the logarithm of some appropriate scale $k$  obeys, with $t=\ln(k/\chi)$,
\ba\label{2}
\partial_t\lambda&=&\beta_\lambda(\lambda,h,g^2),\nonumber\\
\partial_t\gamma&=&A_\mu(\lambda,h,g^2)\gamma.
\ea
Here $h$ stands for Yukawa couplings of $\varphi$ to quarks and leptons - we only keep the dominant coupling of the top quark - and $g$ stands for gauge couplings. The one loop expressions are 
\ba\label{2A}
\beta_\lambda&=&\frac{3}{4\pi^2}(\lambda^2+h^2\lambda-h^4),\\
A_\mu&=&\frac{3}{8\pi^2}(\lambda^2+h^2),\label{3}
\ea
where we omit the contribution from gauge couplings for simplicity of the presentation. Higher loops add terms to $\beta_\lambda$ and $A_\mu$ that involve higher powers of $\lambda$ or $h^2$. They do not change the structure of the flow equations \eqref{2}.

The crucial feature is the vanishing of $\partial_t\gamma=\beta_\gamma$ for $\gamma=0$, such that $\beta_\gamma$ is governed by the ``anomalous mass dimension'' $A_\mu$. This is a consequence of the essentially second order character of the zero-temperature electroweak phase transition - say as a function of $\gamma$. For an exact second order transition the transition point at $\gamma=0$ must be a fixed point, such that $\beta_\gamma(\gamma=0)=0$ \cite{CW1,CW2,CWR}. An exact fixed point would correspond to an additional symmetry, namely ``low-energy dilatation symmetry'', where distances and low-energy fields as $\varphi$ are scaled according to their dimension (including anomalous dimension), while $\chi$ is kept fixed. This enhanced symmetry for $\gamma=0$ singles out this particular point and makes the value $\gamma=0$ ``natural'' in a technical sense \cite{CWR}.

Possible effects which may turn the transition away from second order are due to running couplings. Indeed, the exact Gaussian fixed point for $h^2=g^2=\lambda=\gamma=0$ extends to an approximate fixed point for small $h^2,g^2,\lambda$ and $\gamma=0$. From the point of view of the Gaussian fixed point the couplings $h^2,g^2,\lambda$ are marginal parameters. If their values are small, the running of the couplings is very slow. These effects can be neglected in a good approximation. Perturbatively, $\gamma=0$ remains a partial fixed point, while non-perturbative effects slightly modify the second order character of the phase transition. For couplings in the observed range the largest such effect is a minimal scale of electroweak symmetry breaking induced by quark-antiquark condensates. This is triggered by the running of the strong gauge coupling. For values of $k$ much larger than the characteristic scale of quark-antiquark condensates the latter can be neglected such that $\gamma=0$ is effectively a partial fixed point. 

The partial fixed point for $\gamma=0$ has a simple but striking consequence. If $\gamma(k)$ is small at some scale $k_0$ larger than $\varphi_0$, the flow equations \eqref{2}, \eqref{3} tell us that $\gamma$ will remain small for all scales $k$ below $k_0$ down to the Fermi scale. This statement is not affected by higher order corrections to $A_\mu$ and $\beta_\lambda$. There is no fine-tuning problem order by order in perturbation theory if one expands the flow equation as appropriate for renormalization gauge improved perturbation theory \cite{CW1}. It is sufficient to find an explanation  for a small value of $\gamma$ at some arbitrary scale $k_0$ within the validity of the flow equations \eqref{2}. Thus the solution for the gauge hierarchy problem may be found in the the TeV range or in the range of $10^{18}$ GeV. The requirement of naturalness does not tell us anything about the scale $k$ where the solution is to be found. 

\medskip\noindent
{\em Stability of electroweak phase transition in grand unified or higher dimensional theories.}

Possible ``high energy solutions'' of the gauge hierarchy problem involve momentum scales of the order of $\chi$ where particles beyond the ones of the standard model are supposed to play a role. In this range of scales both $A_\mu$ and $\beta_\lambda$, as well as the beta-functions for the other couplings of the standard model, may differ substantially from the ``low energy flow'' \eqref{2A}, \eqref{3}. For example, it is conceivable that the scale $\chi$ corresponds to the transition from a higher dimensional world to an effective four-dimensional description. In this case $\chi^{-1}$ is a typical length scale for the additional ``internal dimensions'' and an ``infinite number of particles'' can contribute to $A_\mu$ and $\beta_\lambda$ for $k>\chi$. Another setting concerns the extension of the standard model to a grand unified theory at some scale below $M$. 

The zero-temperature electroweak phase transition is of second order also within such an extended short distance theory. Indeed, an exact second order phase transition shows continuity of the order parameter independently of the scale at which one ``looks'' at the theory. The second order character of a phase transition does not depend on the effective degrees of freedom used for a description at a given scale. For example, the short distance theory could involve a large space of couplings. In this case the second order character of the transition implies a hypersurface in the space of couplings for which $\varphi_0$ vanishes. The flow of couplings that are precisely on this hypersurface will remain on the hypersurface. We may now denote by $\gamma$ some characteristic distance (in the space of couplings) from the hypersurface, that will lead to $\varphi_0\neq 0$. The flow of $\gamma$ has to vanish for $\gamma=0$ and will typically be characterized by an anomalous dimension according to eq. \eqref{2}. Only the value of $A_\mu$ will differ between the short distance theory and the standard model. (This argument remains valid as well if the short distance theory has less parameters than the standard model.)

We emphasize in this context that the dilatation symmetry associated to the second order phase transition is easily seen only for an appropriate choice of parameters. Indeed, $\gamma$ should measure a distance from the critical hyperface in coupling constant space which is orthogonal to the couplings parametrizing the hypersurface. Consider some other (very small) coupling $\eta$ not related to this distance. Since the couplings within the hypersurface are allowed to flow, the leading order flow equation for very small $\eta$ is typically $\partial_t\eta=B$. Here $B$ may depend on other couplings parametrizing the hypersurface. For any linear combination, $\alpha=c_1\gamma+c_2\eta, c_i\neq 0$, the beta-function $\beta_\alpha=\partial_t\alpha$ involves a constant term $c_2B$, such that $\alpha=0$ is not stable with respect to the flow. For an inappropriate choice of parameters one could then naively infer additional tuning problems. 

This situation often occurs in grand unified theories where $\gamma$ can be a complicated function of couplings $f_i$ that are specified by other criteria, as multiplying invariants with respect to a grand unified symmetry. We may denote by a set $\{f^c_i\}$ a point on the critical hypersurface. The flow of an individual deviation from the hypersurface, $\delta f_k=f_k-f^c_k$, corresponds, in general, to the flow of the coupling $\alpha$ and does not vanish for $\delta f_k\to 0$. This can mislead to inaccurate claims that the small value of the Fermi scale is unstable if intermediate scales are present. Such an intermediate scale can be associated to some particular $f_k$. A change of the intermediate scale $f^c_k\to f'_k$ can be accompanied by a change in other couplings such that the new value $f'_k$ corresponds again to a point on the critical surface. The parameter $\gamma$ is given by a linear combination $\gamma=\sum_ia_i\delta f_i$ such that the constant terms $B_i$ in the flow of $\delta f_i$ cancel for the flow of $\gamma$. 

\medskip\noindent
{\em High scale attraction.}

Speculations about possible high-energy solutions of the gauge hierarchy problem often invoke a new fixed point. This fixed point corresponds to a vanishing flow of (dimensionless renormalized) couplings for scales $k$ larger then $\chi$. It may be called ``ultraviolet fixed point'', in distinction to the ``approximate infrared fixed point'' for $k\ll\chi$. The flow of couplings can be viewed as a crossover from the vicinity of the ultraviolet fixed point to the infrared fixed point. For any (perturbatively or non-perturbatively) renormalizable theory containing the standard model, the second order character of the electroweak phase transition guarantees the presence of a fixed point both for the ultraviolet and the infrared regime. However, the spectrum of effectively massless excitations typically differs between the ultraviolet and infrared fixed points, such that the anomalous dimension $A_\mu$ for the ultraviolet fixed point differs from eq. \eqref{3}. The ultraviolet fixed point may be used to render a perturbatively non-renormalizable theory non-perturbatively renormalizable. An interesting candidate is the ``asymptotic safety'' scenario for gravity \cite{We}, \cite{Re}.

Let us now suppose that for the ultraviolet fixed point the anomalous mass dimension $A_\mu$ turns out to be large. Then the fast running of $\gamma$ towards small values in the vicinity of the fixed point could lead to a natural solution of the gauge hierarchy problem \cite{CW2}. For constant $A_\mu$ the solution 
\be\label{5AA}
\gamma(k)=\gamma(k_0)\left(\frac{k}{k_0}\right)^{A_\mu}
\ee
could yield $\gamma(\chi)\approx 10^{-32}$ even if one starts with $\gamma(k_0)\approx 1$ for some scale $k_0$ sufficiently above $\chi$. We may call this scenario ``high scale attraction''. In the general language of the renormalization group the deviation from the transition between broken and unbroken electroweak gauge symmetry can be parametrized by a dimensionless parameter $\gamma\chi^2/k^2$. This is a relevant parameter for $k\ll\chi$ where $A_\mu$ is small. For constant $\chi$ it would become irrelevant if $A_\mu>2$ for some new ultraviolet fixed point. On the other hand, for $k\gg \chi$ one often finds a situation where $\chi$ is replaced by a $k$-dependent expectation value $\chi(k)\sim k$. Then $A_\mu>0$ is sufficient to turn the distance from the critical surface to an irrelevant parameter. 

High scale attraction is analogous to the solution of the flatness problem in inflationary cosmology. While the critical density $\Omega=1$ corresponds to an unstable fixed point of the time evolution of a Friedman universe, it is a stable fixed point for inflationary cosmology. The deviation $\Omega-1$ turns from an irrelevant parameter during inflation to a relevant one for the time after inflation. Similarly to inflation, there needs to be an end of high scale attraction. One of the marginal or relevant deviations from the ultraviolet fixed point may generate a mass for some of the particles, e.g. by dimensional transmutation. (This mass scale is associated to $\chi$ in our setting.)

The task for a realization of high scale attraction is to find a fixed point with a sufficiently large $A_\mu$. For models close to the standard model a fixed point with large enough Yukawa coupling $h$ could be a candidate \cite{Ba}, but no solution of this type has been found yet. It is not necessary that the scalar doublet $\varphi$ remains a fundamental field for the description of the ultraviolet fixed point. Interesting candidates for new fixed points have have been found \cite{Gie} for non-perturbatively renormalizable four-fermion interactions, but large values of $A_\mu$ have not been observed. As mentioned above, the ultraviolet fixed point could be associated with asymptotic safety for gravity. For any realistic model of this type it is indeed necessary that the zero-temperature electroweak phase transition is essentially of second order. A reliable computation of the anomalous mass dimension $A_\mu$ in this context would be highly appreciable. Even less is known about the properties of possible ultraviolet fixed points in a higher dimensional setting. We conclude that so far the search for a solution of the gauge hierarchy by high scale attraction has remained inconclusive. 

A fixed point for which $\gamma$ becomes an irrelevant coupling can be associated with the concept of ``self-tuned criticality''. In the language of critical statistical physics this  means that the critical system has no relevant parameter which must be tuned in order to realize criticality. (Deviations from a fixed point are all irrelevant or marginal.) A two-dimensional example is the low temperature phase in the Kosterlitz-Thouless phase transition \cite{KT}. In four dimensions, self-tuned criticality has been observed for a theory with scalar fields coupled to gauge fields \cite{GW}. We see therefore no strong counterindication why high scale attraction could not be realized. 

\medskip\noindent
{\em Prediction of Higgs boson mass.}

Since from purely theoretical considerations we have no indication at what scale $k$ the solution of the gauge hierarchy problem should be found, one may look for hints from experiments. A low scale solution at scales in the TeV range or somewhat higher could lead to a multitude of possible signatures at high energy colliders or for high precision experiments. No such signal has been found up to now. The issue is more complicated for high scale solutions. If the gauge hierarchy finds an explanation at a scale $k\approx \chi$ the standard model may be valid up to the Planck scale. Then no direct or indirect signatures of additional particles beyond the standard model are expected. There is, however, one salient characteristic of high scale attraction, namely that the running of the couplings of the standard model follows the perturbative $\beta$-functions over a very large range of scales. As long as no details of a possible high scale solution are known the only constraints or predictions for the effective low energy theory arise from the running of couplings between the scales $k=\chi$ and $k=\varphi_0$. This typically results in bounds or predictions for the Higgs boson and top-quark masses.

Below the scale $\chi$ eqs. \eqref{2A}, \eqref{3} become valid, together with a similar equation for the running of the top quark Yukawa coupling (omitting for simplicity again contributions from gauge couplings)
\be\label{4}
\partial_t h=\beta_h=\frac{9}{32\pi^2}h^3.
\ee
The system of flow equations \eqref{2A}, \eqref{4} leads to a partial infrared fixed point for the ratio $\lambda/h^2$ \cite{CW2}, \cite{CWR}, \cite{SWa}
\be\label{5}
\left(\frac{\lambda}{h^2}\right)=x_0=(\sqrt{65}-1)/8.
\ee
Indeed, with $x=\lambda/h^2-x_0$ and flow variable $s$ defined by $\partial s/\partial t=h^2$, eqs. \eqref{2A} and \eqref{4} can be combined to
\be\label{7a}
\frac{\partial x}{\partial s}=\frac{3}{4\pi^2}x\left(x+2x_0+\frac14\right)
\ee
and we observe a vanishing flow of $x$ for $x=0$.

However, there is only a finite range of running between $\chi$ and $\varphi_0$ such that the fixed point is not reached precisely. It is rather replaced by an {\em infrared interval} \cite{CWR} with upper and lower bounds $\lambda_{min}$ and $\lambda_{max}$. This  infrared interval is the image of the interval of allowed values of $\lambda$ at the scale $\chi$. Any well defined realistic model requires $\lambda(\chi)$ to be positive - more precisely electroweak symmetry breaking at high scale $\sim\chi$ must be avoided. (Effective potentials with a metastable vacuum and $\varphi_0=175$ GeV seem hard to be realized in a full treatment beyond perturbation theory if the true vacuum has $\varphi_0\equiv 10^{18}$ GeV.) On the other end we only require $\lambda(\chi)<\infty$. The renormalization flow maps $\lambda(\chi)$ to $\lambda(\varphi_0)$ - the interval $[\lambda_{{\rm min}},\lambda_{{\rm max}}]$ being the image of $[0,\infty]$. This renormalization map is highly non-linear. A substantial range of small $\lambda(\chi)$ is mapped to a value very close to $\lambda_{{\rm min}}$, whereas the range of large $\lambda(\chi)$ corresponds to the close vicinity of $\lambda_{{\rm max}}$. (For $\varphi_0/\chi\to 0$ the infrared interval would shrink to one point given by eq. \eqref{5}.)

The infrared interval depends on the top quark mass $m_t=h(\varphi_0)\varphi_0$ in two ways: First, the partial fixed point \eqref{5} as the ``central value'' of the interval involves $h$ and therefore $m_t$. Second, the interval shrinks faster during the renormalization flow for larger $h$ \cite{CWR}. All characteristic features of the infrared interval remain valid in the presence of gauge couplings - only the numerical values of $\lambda_{{\rm min}}$ and $\lambda_{{\rm max}}$ are modified. For a top-quark pole mass of $171$ GeV and including effects from gauge couplings and two loops one finds \cite{SW} for the mass of the Higgs scalar that corresponds to $\lambda_{min}$ and $\lambda_{max}$
\be\label{6}
m_{min}=126 GeV~,~m_{max}=174 GeV.
\ee
Including three loop running and assuming a top quark pole mass of $173$ GeV one finds $m_{{\rm min}}=129~{\rm GeV}$ \cite{AA}, \cite{AB}. The uncertainty of these values amounts to a few GeV. 

Extrapolating the running couplings from the Fermi scale towards shorter distance scales an interval of the type \eqref{6} follows from the requirement of validity of perturbation theory and positiveness of $\lambda$ \cite{z1}, with the concept that new physics has to set in for scales smaller than $\chi$ if $m_H$ is found outside the interval \eqref{6}. In our setting the infrared interval arises as a {\em prediction} of the Higgs boson mass for a scenario of high scale attraction with validity of the standard model up to the Planck mass. (This prediction is not restricted to perturbation theory.) In the following we will argue further that a large class of such models predicts the Higgs boson mass to be very close to the lower bound at $m_{{\rm min}}$. 

Consider a scenario with a high scale fixed point where $\lambda =0$, as advocated in the context of non-perturbative renormalizability of gravity in ref. \cite{SW}. (For such a fixed point also $\gamma$ in eq. \eqref{1} may vanish such that the effective potential could become independent of $\varphi$.) For $k$ below $\chi$ the particles with mass $\sim\chi$, which are supposed to be responsible for the existence of the ultraviolet fixed point, decouple from the flow such that eq. \eqref{2A} becomes valid. Due to the term $\sim-h^4$ the quartic scalar coupling starts to deviate from the fixed point value $\lambda=0$ and increases as $k$ is lowered. It will then be attracted towards the lower bound of the infrared interval, resulting in $\lambda(\varphi_0)=\lambda_{min}$ and $m_H=m_{{\rm min}}$. An experimental finding of $m_H$ near $126$ GeV can be taken as a strong indication in this direction. The scenario remains valid for a high energy fixed point with a small nonzero value of $\lambda$. More generally, the prediction $m_H=m_{{\rm min}}$ results whenever $\lambda(\chi)$ is sufficiently small. A whole range of small quartic couplings at the scale $\chi$ is mapped to $\lambda(\varphi_0)\approx \lambda_{min}$ by the renormalization flow, resulting in a rather robust prediction.

A logical alternative would be a fixed point with large values of $h$ and $\lambda$, as investigated in \cite{CW2}, \cite{Ba}. Large values of $\lambda(\chi)$ are all mapped to the upper bound of the infrared interval and result in $m_H\approx 174$ GeV. This seems to be excluded by the LHC-Higgs bounds. Thus for any scenario with an ultraviolet fixed point a zero or small value $\lambda_*$ seems indicated. The measured value of the Higgs boson mass provides for essential information about the properties of a possible ultraviolet fixed point. 

\medskip\noindent
{\em Prediction of top quark mass.}

The scenario of high scale attraction, together with a transition to the standard model near the Planck scale and $\lambda(\chi)$ close to zero, provides also information about the mass of the top quark. First, on a phenomenological level the identification of the measured Higgs boson mass with $m_{{\rm min}}$ restricts the value of the top quark pole mass. (Recall that $m_{{\rm min}}$ depends on $h$ and therefore on $m_t$.) For $m_{{\rm min}}=126$ GeV one infers a top quark pole mass close to $171.5$ GeV. For a given measured Higgs boson mass this can be taken as a {\em prediction} of the scenario. The presently quoted top quark mass of $173$ GeV is somewhat higher, but uncertainties are a few GeV \cite{AB}. A precise measurement of the top quark pole mass can therefore be used for a possible falsification of our scenario and merits experimental and theoretical effort. 

On a more theoretical level our scenario of high scale attraction entails a lower bound for the top quark mass \cite{SW}. Indeed, for $\lambda(\chi)=0$ one also needs the beta-function $\beta_\lambda$ to be negative or zero at this scale. For positive $\beta_\lambda(\chi)$ the quartic coupling $\lambda(k)$ would turn negative for $k<\chi$, thus inducing high scale electroweak symmetry breaking, in contrast to $\varphi_0=175$ GeV. A negative $\beta_\lambda$ requires the Yukawa couplings $h(\chi)$ to be sufficiently large as compared to the gauge couplings $g(\chi)$. (The gauge couplings make a positive contribution to $\beta_\lambda$.) This provides for a lower bound for the top quark pole mass, given by the condition $\beta_\lambda(\chi)=0$. This lower bound is close to the experimental measured value of $m_t$. On the other hand, a value of $m_t$ substantially above the lower bound would imply a value of $m_{{\rm min}}$ that is larger than the observed value of the Higgs boson mass. An experimental upper bound on $m_H$ translates to a phenomenological upper bound for $m_t$, supplementing the theoretical lower bound. 

A particularly interesting set of ``initial conditions'' at the scale $\chi$ is 
\be\label{9}
\lambda(\chi)=0~,~\beta_\lambda(\chi)=0. 
\ee
The two conditions predict two parameters of the standard model, namely
\be\label{10}
m_H=126~{\rm GeV}~,~m_t=171.5~{\rm GeV},
\ee
where we associate $\chi$ to a scale close to the reduced Planck mass. For an ultraviolet fixed point one has necessarily $\beta_\lambda(k)=0$ for $k\gg\chi$. It is not trivial how this translates to $\beta_\lambda(k)=0$ for a value of $k$ close to $\chi$ where only the particles of the standard model are effectively massless. In principle, the decoupling of heavy particles could lead to a jump of $\beta_\lambda(k)$ between $k\gtrsim \chi$ and $k\lesssim \chi$. The condition $\beta_\lambda(\chi)=0$ amounts therefore to a property of smoothness of the running of $\lambda$ for $k$ larger or smaller than $\chi$. In other words, the particles decoupling at the scale $\chi$ should only give a small contribution to $\beta_\lambda$. (This would be the case if their contribution is $\sim\lambda$, as for the graviton which decouples effectively due to the gravitational interaction becoming very small.)

Of course, the measurement of the two parameters $m_H$ and $m_t$ can only be used for a possible falsification of our scenario, not for a confirmation. One can think of many alternative models where parameters can be adjusted in order to reproduce the correct values for $m_H$ and $m_t$. For example, a similar range for $m_H$ and $m_t$ can be inferred from the requirement that the Higgs potential should have two minima \cite{FN} or from speculations about vanishing quadratic divergences \cite{20St}. It seems worthwhile to reduce the uncertainties, both for the measured values of the Higgs boson mass and the top quark pole mass. On the theoretical side one may explore possible modifications from ``intermediate scales'' that only mildly influence the running of couplings. This concerns a possible grand unification at a scale only somewhat below the Planck mass, as well as effects of an intermediate scale related to $B-L$ violation that governs the size of the neutrino masses. 

We conclude that experimental hints towards a high-scale solution of the gauge hierarchy problem are necessarily much weaker than the possibilities of direct or indirect detection of new particles for a low-scale solution. Nevertheless, an agreement of the observed Higgs boson and top quark masses with the lower bound of the infrared interval may point towards a high scale solution with a small value of the quartic scalar coupling at the unification scale, possibly  zero and corresponding to a fixed point. It seems worthwhile to remain open minded about the scale where the gauge hierarchy problem may be solved and to devote an increased theoretical effort into ideas for high scale attraction.

\end{document}